\documentclass[11pt]{article}

\usepackage{EMNLP2023}

\usepackage{times}
\usepackage{latexsym}
\usepackage{mathtools}

\usepackage[T1]{fontenc}

\usepackage[utf8]{inputenc}

\usepackage{microtype}

\usepackage{inconsolata}
\usepackage{threeparttable}
\usepackage[ruled,linesnumbered]{algorithm2e}

\usepackage{amssymb,bm}
\usepackage{amsmath}

\usepackage{pifont}
\usepackage{color,xcolor}
\usepackage{multirow}
\usepackage{listings}
\usepackage[shortlabels]{enumitem}
\usepackage{tcolorbox} 
\usepackage{url}
\usepackage{subcaption}

\newcommand{\F}{Figure}

\newcommand{\T}{Table}
\renewcommand{\S}{Sec.}

\newcommand{\ignore}[1]{}

\newcommand{\parh}[1]{\noindent\textbf{#1}}

\newcommand{\tool}{\textit{InsightPilot}\xspace}

\definecolor{pptred}{RGB}{176,36,24}

\title{\tool: An LLM-Empowered Automated Data Exploration System}

\author{Pingchuan Ma \\ HKUST \And
        Rui Ding$^\dagger$ \\ Microsoft Research \And
        {\bf Shuai Wang}$^\dagger$ \\ HKUST \\
        \AND 
        Shi Han \\ Microsoft Research \And
        Dongmei Zhang \\ Microsoft Research}

\begin{document}
\maketitle
\begin{abstract}

Exploring data is crucial in data analysis, as it helps users understand and
interpret the data more effectively. However, performing effective data
exploration requires in-depth knowledge of the dataset, the user intent and
expertise in data analysis techniques. Not being familiar with either can create
obstacles that make the process time-consuming and overwhelming.

To address this issue, we introduce \tool, an LLM (Large Language Model)-based,
automated data exploration system designed to simplify the data exploration
process. \tool features a set of carefully designed analysis actions that
streamline the data exploration process. Given a natural language question,
\tool collaborates with the LLM to issue a sequence of analysis actions, explore
the data and generate insights. We demonstrate the effectiveness of \tool\ in a
user study and a case study, showing how it can help users gain valuable
insights from their datasets.

\end{abstract}

\section{Introduction}

Exploratory data analysis (EDA) is a demanding task that extracts meaningful
insights from data~\cite{Komorowski2016, jebb2017exploratory, devore2007making}.
Data exploration is a critical step in data analysis. In general, it involves a
series of data analysis operations, such as filtering, sorting, and grouping, to
discover patterns in data. Usually, the process is iterative and interactive,
and the user needs to manually explore the data back-and-forth to gain insights.
This process is often time-consuming and requires considerable domain knowledge
and expertise. Below, we present an example to illustrate a data
exploration process.

\begin{figure*}[!htbp]
    \centering
    \includegraphics[width=0.99\linewidth]{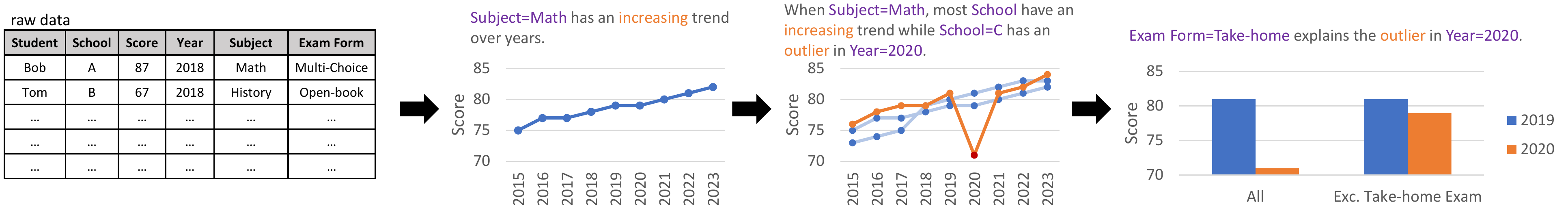}
    \vspace{-8pt}
    \caption{An example of data exploration.}
    \vspace{-15pt}
    \label{fig:example}
\end{figure*}

\parh{Example.}~\textit{Using a student performance dataset from multiple
schools (\F~\ref{fig:example}), an education analyst, Alice, conducts EDA to
comprehend trends in math performance. After considerable manual data filtering
and sorting, Alice captures an upward trend by plotting math scores over time.
Alice then puts in more effort into manual data filtering for comparing student
performance across schools A, B, and C. Finally, she observes that both schools
A and B illustrate an increasing trend while school C has an outlier in 2020.
Alice is curious and decides to investigate the outlier. She spends even more
time exploring the data back-and-forth, filtering and grouping by various
variables until she finally finds that when excluding ``take-home'' exams, the
outlier for school C in 2020 is no longer present. Alice notes this finding and
concludes that the outlier is caused by a policy change of the exam form in
school C in 2020.}

Alice's manual data sifting for insights is effort-intensive and time-consuming,
highlighting the need for an efficient automated data exploration system to
simplify the process.

\parh{Existing Solutions.}~To date, a number of data exploration systems have
been proposed in the data management and data mining
community~\cite{bar2020automatically, chanson2022automatic,
personnaz2021balancing, cao2023learn}. In general, these systems leverage a
heuristic score function to identify the ``best'' data exploration sequence (a
series of data analysis operations). While these systems show potential, they
exhibit key limitations. \ding{202} \textbf{User Intent Ignorance}: Existing
tools are designed for general exploration and often fail to incorporate user
intent. For instance, an analyst may be interested in understanding the
economics-related factors but receive insights about demographics. \ding{203}
\textbf{Dataset Characteristic Ignorance}: They overlook dataset
characteristics, often providing irrelevant insights. For instance, in a flight
delay dataset, a correlation between flight delays and weather might yield
insights but irrelevant factors like time and weather does not make sense in the
context. They fall short in delivering a direct answers to the user question.

Recently, large language models (LLM) have shown promising potential in
understanding user intent and generating actions to achieve user-specified
goals~\cite{yao2022react}. In this regard, we anticipate that LLM can be
leveraged to drive the data exploration process. However, there are several
challenges that impede the adoption. \ding{204} \textbf{Hallucination}: Due to
the infamous hallucination issue~\cite{ji2023survey}, LLMs often generate
unreliable contents and are thus not mature for production use. \ding{205} 
\textbf{Overwhelming Context Window}: A dataset may contain millions of 
cells, which is overwhelming for LLMs to process.

\parh{Our Solution.}~To address these challenges, we propose \tool, a system
that automates data exploration using LLMs. This system facilitates exploration
through the synergy of an LLM and an insight engine, which integrates three
production-quality insight discovery tools:
QuickInsight~\cite{ding2019quickinsights}, MetaInsight~\cite{ma2021metainsight},
and XInsight~\cite{ma2023xinsight} (detailed in \S~\ref{subsec:iquery}). These
tools offers a unified insight representation, enabling the LLM to engage
coherently. The insight engine provides the LLM with accurate and reliable
insights, avoiding hallucination. Furthermore, the insight engine presents a
concise abstraction of the dataset to alleviate the overwhelming context window
issue. In \tool, users input high-level queries, like ``\textit{show me the
interesting trend in mathematics scores for students}''. Then, the \tool employ
an LLM to interact with the insight engine using a set of carefully designed
analysis actions to streamline common data exploration tasks. These actions
serve as a coherent transition to chain up insights and generate a data
exploration sequence to answer the user's question. Finally, \tool\ summarizes
the results using natural language together with charts that are understandable
to non-technical users.

\parh{Contributions.}~In summary, we make the following contributions: We
propose \tool, an automated system for data exploration that employs LLMs to
drive the exploration process. \tool\ streamlines the exploration process by
interacting with an insight engine using a set of carefully designed analysis
actions. We conduct a user study and a case study to demonstrate the
effectiveness of \tool\ in real-world scenarios.

\section{Related Work}

\parh{Text-to-SQL.}~To date, text-to-SQL is the most popular approach to
enabling natural language interface to database. It translates users' utterances
into SQL queries for relational databases and has been studied by both database
and NLP communities for several decades~\cite{yu2018spider,
kim2020natural,ma2022mt}. Recent studies have shown that with LLM, text-to-SQL
can now be augmented to support non-SQL enquiries such as entity
extraction~\cite{cheng2022binding}. However, in EDA, users' intents are often
more complex than simple SQL queries. EDA generally involves more complicated
user intents and goes beyond the expressiveness of basic SQL queries. We take a
step further by using an LLM and analysis actions in \tool, to produce natural
and coherent data exploration sequences that accurately address users'
questions. This innovation provides an important complement to the existing
text-to-SQL approach.

\parh{Analytics Model in OLAP.}~Traditionally, users interact with OLAP (online
analytical processing) systems with a set of pre-defined operators (e.g.,
drill-down and roll-up)~\cite{vassiliadis1999survey}. Recently, there is a surge
of interest in developing analytics models with higher-level abstraction and
automation to facilitate complex OLAP needs~\cite{vassiliadis2019beyond}. In
\tool, we use ``\textit{analysis actions}'' to describe such high-level
abstractions.

\section{Preliminaries}
\label{sec:prel}
In this section, we introduce the preliminaries of exploratory data analysis.

\parh{Data Model.}~Let $D\coloneqq\{X_1, \cdots,X_n\}$ represents
multi-dimensional data comprising $n$ attributes, where each attribute $X_i$ is
either a dimension or a measure. A \textbf{dimension} $X_i$ is a categorical
attribute that can be used to group data. A \textbf{measure} $X_i$ is a
numerical attribute that can be used to perform aggregation operations. In
\tool, \textbf{filter} is the basic unit of data operations. Given a
multi-dimensional data $D$ and a dimension $X$, a filter $p_i={X=x_i}$ (e.g.,
``Subject=Math'') implies an equality assertion to $X$ such that the value of
$X$ will equal $x_i$. A \textbf{subspace} is a conjunction of filters on
disjoint dimensions (e.g., ``Subject = Math AND Year = 2019''). A
\textbf{breakdown} dimension is the dimension where the group-by operation is
performed. Given a measure $M$, users may perform aggregation operations (such
as \texttt{SUM} and \texttt{AVG} in SQL) over some records for $M$.

\parh{Analysis Entity (AE).}~An AE is defined as a 3-tuple $\textit{AE}\coloneqq
\langle \textit{agg}(M), S, B\rangle$, where $M$ is a measure with an
aggregation function $\textit{agg}$ applied, $S$ is a subspace, and $B$ is a
breakdown dimension. It can be interpreted as an equivalent SQL query that
performs aggregation operations over a set of records for the measure $M$ in a
subspace $S$, grouped by the breakdown dimension $B$. For instance, the AE
$\langle \texttt{AVG(Score)}, \texttt{Subject = Math}, \texttt{Year}\rangle$ is
equivalent to the SQL query \textcolor{blue}{\small\texttt{SELECT AVG(Score)
FROM Table WHERE Subject = Math GROUP BY Year}}.


\parh{Data Insight.}~A basic data insight is represented as a 3-tuple $\langle
\textit{AE}, \textit{Type}, \textit{Property}\rangle$. Here, $\textit{AE}$
denotes an analysis entity, $\textit{Type}$ specifies the insight's kind (e.g.,
trend, outlier), and $\textit{Property}$ encapsulates additional outputs from
insight mining algorithms, like extreme points for a unimodality pattern. We
categorize insights into \textbf{basic insights}, directly derived from data
(e.g., trend insights), and \textbf{compound insights}, which build upon other
insights. A \textit{meta-insight}, for instance, summarizes several similar
insights (e.g., sales trends across various cities). Both insight categories can
be expressed as the 3-tuple format. Throughout this paper, "insight" pertains to
both types. We have crafted templates to articulate these insights in
user-friendly language, accompanied by visualizations.

\begin{figure*}[!htbp]
    \centering
    \includegraphics[width=0.99\linewidth]{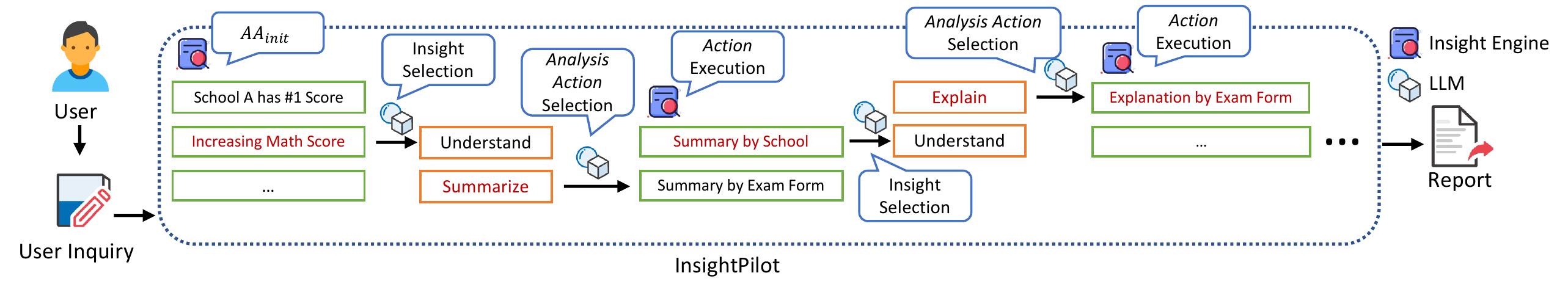}
    \vspace{-15pt}
    \caption{Pipeline of \tool. Selected insights/analysis actions are
    highlighted in \textcolor{pptred}{red}.}
    \vspace{-15pt}
    \label{fig:pipeline}
\end{figure*}

\section{Problem Definition}
\label{sec:problem}
In this section, we define the problem of generating a sequence of data insights
to address users' analysis intents.

\parh{Analysis Actions.}~An analysis action is defined as a transition from one
data insight to several other data insights (which may also be no insight or
solely one insight). In our context, an action represents a reasonable data
analysis operation with the existing knowledge (e.g., user question, dataset,
and explored insights). This is defined as a function $\textit{AA}:
\textit{Insight} \rightarrow \textit{Insight}^*$, where $\textit{Insight}^*$
denotes the set of all possible data insights. In addition, we define two
special analysis actions, namely, $\textit{AA}_{\textit{init}}$ and
$\textit{AA}_{\textit{back}}$. $\textit{AA}_{\textit{init}}$ that takes a
dataset as input and provides a set of initial insights and
$\textit{AA}_{\textit{back}}$ that backtracks to the last state and returns the
insights generated by the preceding analysis action.

\parh{Data Exploration Sequence.}~A data exploration sequence is defined as a
series of data insights interconnected by analysis actions. It is represented as
$\mathcal{S}=\langle \textit{AA}_{\textit{init}}, \textit{Insight}_1,
\textit{AA}_1, \cdots, \textit{AA}_n, \bot\rangle$, where $\textit{Insight}_i$
is a data insight picked from the output of the preceding analysis action
$\textit{AA}_{i-1}$, $\textit{AA}_i$ is an analysis action, and $\bot$
symbolizes the termination. Each analysis action $\textit{AA}_i$ takes the
preceding data insight $\textit{Insight}_{i-1}$ as input and produces multiple
insights to be picked for the next analysis action. 

\parh{Generating Final Answer.}~Given a user question $Q$ and the data
exploration sequence $\mathcal{S}$, we define the problem of generating a final
answer as a function $\textit{FA}: Q \times \textit{flatten}(\mathcal{S})
\rightarrow A$, where $A$ is the final answer to the user question $Q$ and
$\textit{flatten}(\mathcal{S})$ is the top-k of all data insights generated in
the data exploration sequence (including unpicked insights). The final answer
$A$ can be obtained by any document QA techniques over
$\textit{flatten}(\mathcal{S})$.

\parh{Application Scope.}~In \tool, we focus on the class of data analysis tasks
are expressed as \textit{fuzzy} and \textit{high-level} tasks. These are tasks
where the user's intent is not explicitly clear or the analysis objective is
often complex, requiring multiple steps to fully address. For example, a fuzzy
high-level task could be ``\textit{Analyze sales performance over the past
year.}'' The exact steps required to answer this question are not specified, and
a variety of different analysis actions and insights may be required to provide
a comprehensive answer. Note that while we focus on this specific class of
questions, the unique feature of \tool can be integrated with other systems
(e.g., text-to-SQL) and support diverse data analysis scenarios.
\section{\tool\ Design}

\F~\ref{fig:pipeline} depicts the overview of \tool. Overall, \tool\ constitutes
a pipeline of three components: (1) a user interface that enables users to issue
inquiries in natural language, and also depicts analysis results in texts and
charts; (2) a LLM that drives the exploration process by selecting appropriate
insights and analysis intents based on the context (e.g., user question, dataset
domain knowledge, and current exploration state); (3) an insight engine that
executes analysis action, generates insights, and presents results in natural
language. 

\parh{Working Example.}~In \F~\ref{fig:pipeline}, \tool is illustrated using the
example from \F~\ref{fig:example}. A user poses a query: ``\textit{show me any
interesting trend in mathematics scores for students}''. The insight engine then
generates initial insights with $\textit{AA}_{\textit{init}}$. One such insight
might be ``\textit{School A has the Rank\#1 average score.}'' Based on the
user's question, the LLM identifies the most pertinent insight, such as
``\textit{the mathematics scores of students have been increasing over time}'',
using predefined prompts (refer \S~\ref{subsec:prompt}). After choosing the
insight, we prepare potential analysis actions and the LLM selects an
appropriate one, in this case, \textit{compare} (details in
\S~\ref{subsec:iquery}). Executing this action, the insight engine summarizes
the math scores trend across schools. It observes: ``\textit{most schools show
rising math scores, except for an outlier in 2020 for school C.}'' To delve
deeper, the LLM continues to select insights and actions, eventually querying
the engine to ``\textit{explain the 2020 outlier for school C}''.

Interactions continue until the LLM completes its exploration (i.e., choose
$\bot$ as the next action) or hits the token size limit. Once done, insights are
translated to natural language for the prompt. Given the typically large number
of insights, insight ranking becomes crucial. The insight engine then presents
the top-$K$ insights, which the LLM condenses into a coherent report. This
report and the top-$K$ insights (in the form of charts) are then displayed to
the user via the interface.

\subsection{Analysis Action}
\label{subsec:iquery}

Every time the LLM selects an insight and an analysis action, the insight engine
will execute the action and generate new insights. Currently, we have prepared
four analysis actions for the LLM to select: \textit{understand},
\textit{summarize}, \textit{compare}, and \textit{explain}. These actions are in
accordance with three insight discovery solutions, namely
QuickInsight~\cite{ding2019quickinsights} for \textit{understand},
MetaInsight~\cite{ma2021metainsight} for \textit{summarize} and
\textit{compare}, and XInsight~\cite{ma2023xinsight} for \textit{explain}. We
now elaborate on the design of these analysis actions.

\parh{\ding{192} Understand.}~This action is designed to help users understand
the high-level patterns in the data. In particular, it attempts to enumerate all
possible AEs (see definition in \S~\ref{sec:prel}) under the AE of input
insight, applies the insight mining algorithm (e.g., trend detection) on each AE
to identify basic insight and transforms them into human-understandable natural
language.

\parh{\ding{193} Summarize.}~This action aims to view an input insight from
various angles. Starting with a basic insight (like a trend), it employs a
specific insight mining algorithm on the AEs of the input to verify the presence
of the primary insight type and property (e.g., an increasing trend) across each
AE. If consistent across all AEs, the output is ``\textit{the basic insight type
and property are universal among AEs}''. If not, it's ``\textit{the basic
insight type and property are present in some AEs}''. Using the example insight
of an ``\textit{increasing math score trend}'', the outcome could be
``\textit{most schools show rising math scores, barring school C}'' or
``\textit{in most subjects, scores have risen over time}''. These compound
insights can be further explored by subsequent analysis actions.

\parh{\ding{194} Compare.}~This action shares a similar design with
\textit{summarize}. It is designed to compare the input insight from different
neighbors. In particular, it starts with a basic insight (e.g., a trend) and
then applies the particular insight mining algorithm on the neighboring AEs.
Given an input insight ``\textit{the increasing trend of the mathematics scores
in school A}'', it will generate a comparison of the trend regarding different
schools, e.g., ``\textit{the mathematics scores of students in school A and B
have been increasing over time, except school C.}'' This comparison is also
represented by a compound insight.

\parh{\ding{195} Explain.}~This action is designed to explain the insight that
reveals a difference or an outlier in the data. It supports both basic insights
(e.g., a outlier insight or a change point insight) and compound insights (e.g.,
a summary insight with an exceptional case). Given the input insight with
difference or outlier, it will identify a subspace that is responsible for the
outcome using causal inference and then constitute a new compound insight to
encode the cause. For example, given ``\textit{the outlier of the mathematics
scores in 2020 for school C}'', it will identify explanations such as
``\textit{the difference on the mathematics scores of students in school C
between 2019 and 2020 is caused by Exam Form=Take-home. When excluding Exam
Form=Take-home, 2020 is no longer an outlier.}'' 

\subsection{Prompt Engineering}
\label{subsec:prompt}

To deliver a self-contained presentation, we describe the design of our prompt
engineering techniques. In general, any agent-based prompt template (e.g.,
ReACT~\cite{yao2022react}) can be used to instantiate \tool. We prepare separate
prompt templates for different stages of \tool for selecting insights and
analysis actions, and for finalizing the answer. Routine instructions are used
in the prompt to improve the usefulness, clarity, and coherence of the LLM
outputs.

\subsection{Insight Ranking}

Consider the final answer generation phase detailed in \S~\ref{sec:problem}. The
insight engine often yields an overwhelming number of insights, sometimes
reaching hundreds within a single exploration sequence. Given the LLM's
capacity, it is infeasible to process all these insights. Thus, we prioritize by
extracting the top-$K$ insights. Notably, the value of $K$ surpasses the count
of selected insights in the exploration sequence ($\textit{Insight}\in
\mathcal{S}$), ensuring the chosen insights are encompassed within the top-$K$.
Next, we present three schemes to rank top-$K$ insights.

\parh{Redundant Insight Elimination.}~Trivial insights can be eliminated if they
are entailed by a more informative insight. For example, if three insight \#1:
``\textit{School=A has the highest mathematics score,}'' \#2: ``\textit{School=A
has the highest score in 2022,}'' and \#3 ``\textit{School=A has the highest
mathematics score in 2022}'' are generated, we can exclude the third insight
since it trivially derives from the first two. Enlightened by this observation,
we propose to eliminate insights that are entailed by other insights. In
particular, such elimination is achieved by identifying insights by looping
through every possible pair of insights and checking if there exists a dominator
to make one of them trivial.

\begin{figure*}[t]
    \centering
    \includegraphics[width=0.88\linewidth]{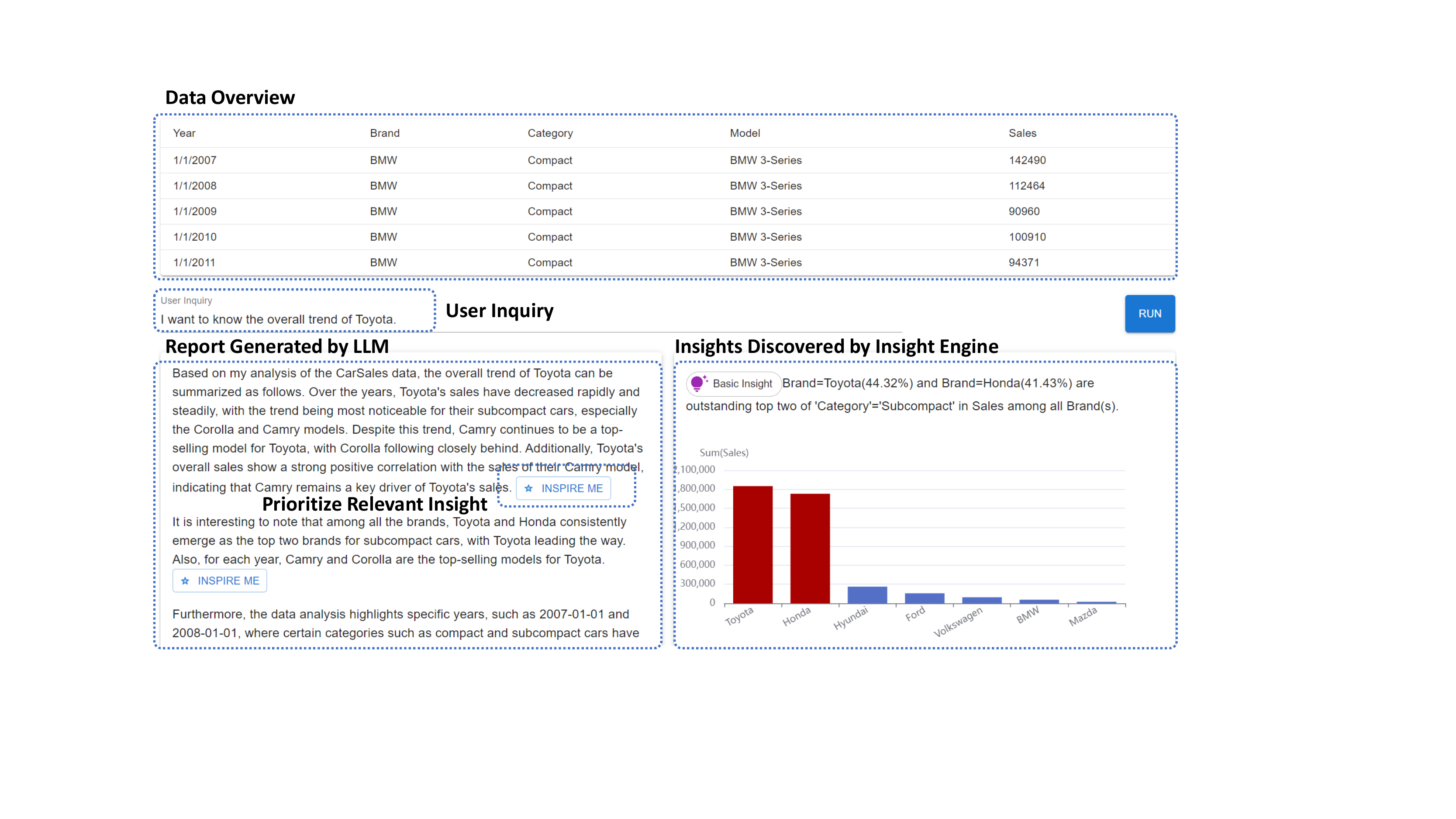}
    \vspace{-5pt}
    \caption{User interface of \tool.}
    \vspace{-8pt}
    \label{fig:demo}
\end{figure*}

\parh{Semantic Similarity-based Elimination.}~To further decrease the number of
insights for the LLM to handle, we employ semantic similarity. We identify the
top-$K'$ ($K'\gg K$) insights most relevant to the user's question using an
embedding model that transforms text inputs into vector representations. By
calculating the cosine similarity between each insight's vector and the user's
question, we rank the insights, selecting the top-$K'$ most relevant ones. This
method not only reduces the LLM's processing load, but it also ensures the
retained insights align closely with the user's question.

\parh{Diversity-aware Reranking.}~After applying the above two strategies, we
obtain the top-$K'$ insights. Then, we seek to re-rank them according to their
diversity to provide the top-$K$ insight. In the context of recommending a set
of insights, the goal is to select insights with high individual scores and low
redundancy. This can be thought of as maximizing the total usefulness of the
selected insights. To achieve this, we leverage the second-order approximated
ranking algorithm as explained in~\cite{ma2021metainsight} to determine the
order.

\section{Evaluation}

\parh{Implementation.}~We implement our tool based on the codebases of
QuickInsight, MetaInsight, and XInsight, adding an additional 1.8K lines of C\#
code and 1.5K lines of JavaScript code. We use ``gpt-3.5-turbo'' as our language
model and ``text-embedding-ada-002'' is used to generate embeddings. Both models
are provided by OpenAI.

\begin{table}[h]
\centering
\begin{threeparttable}
    \resizebox{\linewidth}{!}{
\begin{tabular}{|l|c|c|c|}
\hline
 &  \tool & Code Interpreter & Pandas Agent \\
\hline
\textbf{Relevance} & {\bf 4.50$\pm$0.76} & 4.08$\pm$0.86 & 1.92$\pm$1.00 \\
\hline
\textbf{Completeness} & {\bf 4.67$\pm$0.55} & 3.54$\pm$1.00 & 1.12$\pm$0.33 \\
\hline
\textbf{Understandability} & {\bf 4.46$\pm$0.64} & 4.25$\pm$0.83 & 1.62$\pm$0.90 \\
\hline
\end{tabular}}
\caption{Results of the User Study}
\label{tab:user_study}
\end{threeparttable}

\end{table}

\paragraph{User Study.}~We conduct a user study to simulate the real-world
application of \tool, highlighting its unique advantages over existing solutions
like OpenAI Code Interpreter~\cite{openaicodeinter} and Langchain Pandas
Agent~\cite{langchain}, both being \textit{state-of-the-art} in their domains.
We explored but excluded text-to-SQL models like Flan-T5~\cite{chung2022scaling}
and Anthropic Claude~\cite{claude}, due to their inability to provide direct
answers or load the datasets. Four independent data science participants are
recruited for the study. They are given two datasets and asked to raise three
questions each within the \tool application scope, resulting in 24 groups of
comparisons (4 participants $\times$ 2 datasets $\times$ 3 questions). They
score the systems on \textit{Relevance}, \textit{Completeness} and
\textit{Understandability} (scale of 1 to 5).

Results are reported in \T~\ref{tab:user_study}. \tool consistently outperforms
the others in all three metrics, showcasing its capability in offering relevant,
complete and understandable responses. Specifically, \tool is notably better
than the Code Interpreter and Pandas Agent in \textit{completeness} (p-value $<$
0.05). Upon examining the competitors' responses, we found they often provide an
ad-hoc answer to a specific region of the dataset. For example, when inquiring
about differences in car sales between Mazda and Toyota, competitors reveal only
the overall difference, whereas \tool further analyzes various breakdowns,
identifying that ``\textit{Toyota's Corolla model accounts for a larger
percentage of sales compared to Mazda's models.}'' 

\parh{Case Study.}~We demonstrate \tool's use in \F~\ref{fig:demo}, showcasing a
portion of its output (due to space limit). The user is using a car sales
dataset and enquiries ``\textit{I want to know the overall trend of Toyota.}''
\tool\ first identifies that ``Toyota has a decreasing trend on its sales over
the years'' and then dives into two representative models of Toyota, namely
``Toyota Corolla'' and ``Toyota Camry.'' It identifies that ``Corolla'' and
``Camry'' constitute two top-selling models of Toyota and the sales of ``Camry''
has a strong correlation with the overall Sales of Toyota. Therefore, \tool\
concludes that the Camry is the key driver of Toyota's sales. Afterwards, \tool\
further compares Toyota with Honda and identifies that they are the top-two
brands for subcompact cars while Toyota leading the way. Besides, \tool\ also
looks into the sales of Toyota in different years and obtains other interesting
insights.

\section{Discussion}

\parh{Action-wise Performance.}~The efficacy of \tool, an automated data
analytics tool, hinges on the comprehensive design of each action and its
accurate execution. While we introduce four actions rooted in common data
analysis techniques, it is vital to note that \tool's innovation is not tied to
specific action designs. Instead, its uniqueness lies in leveraging LLM for data
exploration.

\parh{Comprehensive Assessment.}~To validate \tool's effectiveness, it is
essential to evaluate it across diverse real-life datasets and dimensions, such
as keyword preservation. Nonetheless, \tool often produces open-ended responses,
making manual evaluation crucial for assessing answer quality. These hurdles
make it challenging to efficiently evaluate \tool's performance in a
comprehensive manner. An LLM-based evaluation framework could potentially
streamline this process~\cite{wang2023large, li2023split}.

\section{Conclusion}

In this paper, we introduce \tool, an LLM-empowered automated data exploration
system. By seamlessly integrating LLM with state-of-the-art insight engines,
\tool streamlines data analysis into a coherent exploration sequence. Effective
for real-world datasets, it allows users to derive insights via natural language
inquiries. \tool equips even non-technical individuals to benefit from data
analysis, bolstering efficiency and data-driven decision-making.

\section*{Acknowledgement}

Rui Ding and Shuai Wang are the corresponding authors.

\bibliography{main}

\begin{thebibliography}{24}
\expandafter\ifx\csname natexlab\endcsname\relax\def\natexlab#1{#1}\fi

\bibitem[{Anthropic(2023)}]{claude}
Anthropic. 2023.
\newblock Anthropic claude chat.
\newblock \url{https://claude.ai/chat/}.

\bibitem[{Bar~El et~al.(2020)Bar~El, Milo, and Somech}]{bar2020automatically}
Ori Bar~El, Tova Milo, and Amit Somech. 2020.
\newblock Automatically generating data exploration sessions using deep reinforcement learning.
\newblock In \emph{Proceedings of the 2020 ACM SIGMOD International Conference on Management of Data}, pages 1527--1537.

\bibitem[{Cao et~al.(2023)Cao, Xie, and Huang}]{cao2023learn}
Yukun Cao, Xike Xie, and Kexin Huang. 2023.
\newblock Learn to explore: on bootstrapping interactive data exploration with meta-learning.
\newblock In \emph{2023 IEEE 39th International Conference on Data Engineering (ICDE)}, pages 1720--1733. IEEE.

\bibitem[{Chanson et~al.(2022)Chanson, Labroche, Marcel, Rizzi, and t'Kindt}]{chanson2022automatic}
Alexandre Chanson, Nicolas Labroche, Patrick Marcel, Stefano Rizzi, and Vincent t'Kindt. 2022.
\newblock Automatic generation of comparison notebooks for interactive data exploration.
\newblock In \emph{EDBT}, pages 2--274.

\bibitem[{Cheng et~al.(2023)Cheng, Xie, Shi, Li, Nadkarni, Hu, Xiong, Radev, Ostendorf, Zettlemoyer et~al.}]{cheng2022binding}
Zhoujun Cheng, Tianbao Xie, Peng Shi, Chengzu Li, Rahul Nadkarni, Yushi Hu, Caiming Xiong, Dragomir Radev, Mari Ostendorf, Luke Zettlemoyer, et~al. 2023.
\newblock Binding language models in symbolic languages.
\newblock \emph{International Conference on Learning Representations}.

\bibitem[{Chung et~al.(2022)Chung, Hou, Longpre, Zoph, Tay, Fedus, Li, Wang, Dehghani, Brahma et~al.}]{chung2022scaling}
Hyung~Won Chung, Le~Hou, Shayne Longpre, Barret Zoph, Yi~Tay, William Fedus, Eric Li, Xuezhi Wang, Mostafa Dehghani, Siddhartha Brahma, et~al. 2022.
\newblock Scaling instruction-finetuned language models.
\newblock \emph{arXiv preprint arXiv:2210.11416}.

\bibitem[{Devore(2007)}]{devore2007making}
Jay Devore. 2007.
\newblock Making sense of data: A practical guide to exploratory data analysis and data mining.

\bibitem[{Ding et~al.(2019)Ding, Han, Xu, Zhang, and Zhang}]{ding2019quickinsights}
Rui Ding, Shi Han, Yong Xu, Haidong Zhang, and Dongmei Zhang. 2019.
\newblock Quickinsights: Quick and automatic discovery of insights from multi-dimensional data.
\newblock In \emph{ACM SIGMOD International Conference on Management of Data}.

\bibitem[{Jebb et~al.(2017)Jebb, Parrigon, and Woo}]{jebb2017exploratory}
Andrew~T Jebb, Scott Parrigon, and Sang~Eun Woo. 2017.
\newblock Exploratory data analysis as a foundation of inductive research.
\newblock \emph{Human Resource Management Review}, 27(2):265--276.

\bibitem[{Ji et~al.(2023)Ji, Lee, Frieske, Yu, Su, Xu, Ishii, Bang, Madotto, and Fung}]{ji2023survey}
Ziwei Ji, Nayeon Lee, Rita Frieske, Tiezheng Yu, Dan Su, Yan Xu, Etsuko Ishii, Ye~Jin Bang, Andrea Madotto, and Pascale Fung. 2023.
\newblock Survey of hallucination in natural language generation.
\newblock \emph{ACM Computing Surveys}, 55(12):1--38.

\bibitem[{Kim et~al.(2020)Kim, So, Han, and Lee}]{kim2020natural}
Hyeonji Kim, Byeong-Hoon So, Wook-Shin Han, and Hongrae Lee. 2020.
\newblock Natural language to sql: Where are we today?
\newblock \emph{VLDB}.

\bibitem[{Komorowski et~al.(2016)Komorowski, Marshall, Salciccioli, and Crutain}]{Komorowski2016}
Matthieu Komorowski, Dominic~C. Marshall, Justin~D. Salciccioli, and Yves Crutain. 2016.
\newblock \href {https://doi.org/10.1007/978-3-319-43742-2_15} {\emph{Exploratory Data Analysis}}, pages 185--203. Springer International Publishing, Cham.

\bibitem[{Langchain(2023)}]{langchain}
Langchain. 2023.
\newblock Langchain pandas dataframe agent.
\newblock \url{https://python.langchain.com/docs/integrations/toolkits/pandas}.

\bibitem[{Li et~al.(2023)Li, Wang, Ma, Wu, Li, Wang, Gao, and Liu}]{li2023split}
Zongjie Li, Chaozheng Wang, Pingchuan Ma, Daoyuan Wu, Tianxiang Li, Shuai Wang, Cuiyun Gao, and Yang Liu. 2023.
\newblock Split and merge: Aligning position biases in large language model based evaluators.
\newblock \emph{arXiv preprint arXiv:2310.01432}.

\bibitem[{Ma et~al.(2021)Ma, Ding, Han, and Zhang}]{ma2021metainsight}
Pingchuan Ma, Rui Ding, Shi Han, and Dongmei Zhang. 2021.
\newblock Metainsight: Automatic discovery of structured knowledge for exploratory data analysis.
\newblock In \emph{ACM SIGMOD International Conference on Management of Data}.

\bibitem[{Ma et~al.(2023)Ma, Ding, Wang, Han, and Zhang}]{ma2023xinsight}
Pingchuan Ma, Rui Ding, Shuai Wang, Shi Han, and Dongmei Zhang. 2023.
\newblock Xinsight: explainable data analysis through the lens of causality.
\newblock In \emph{ACM SIGMOD International Conference on Management of Data}.

\bibitem[{Ma and Wang(2022)}]{ma2022mt}
Pingchuan Ma and Shuai Wang. 2022.
\newblock Mt-teql: Evaluating and augmenting neural nlidb on real-world linguistic and schema variations.
\newblock \emph{VLDB}.

\bibitem[{OpenAI(2023)}]{openaicodeinter}
OpenAI. 2023.
\newblock Openai code interpreter.
\newblock \url{https://chat.openai.com/?model=gpt-4-code-interpreter}.

\bibitem[{Personnaz et~al.(2021)Personnaz, Amer-Yahia, Berti-Equille, Fabricius, and Subramanian}]{personnaz2021balancing}
Aur{\'e}lien Personnaz, Sihem Amer-Yahia, Laure Berti-Equille, Maximilian Fabricius, and Srividya Subramanian. 2021.
\newblock Balancing familiarity and curiosity in data exploration with deep reinforcement learning.
\newblock In \emph{Fourth Workshop in Exploiting AI Techniques for Data Management}, pages 16--23.

\bibitem[{Vassiliadis et~al.(2019)Vassiliadis, Marcel, and Rizzi}]{vassiliadis2019beyond}
Panos Vassiliadis, Patrick Marcel, and Stefano Rizzi. 2019.
\newblock Beyond roll-up’s and drill-down’s: An intentional analytics model to reinvent olap.
\newblock \emph{Information Systems}, 85:68--91.

\bibitem[{Vassiliadis and Sellis(1999)}]{vassiliadis1999survey}
Panos Vassiliadis and Timos Sellis. 1999.
\newblock A survey of logical models for olap databases.
\newblock \emph{ACM Sigmod Record}, 28(4):64--69.

\bibitem[{Wang et~al.(2023)Wang, Li, Chen, Zhu, Lin, Cao, Liu, Liu, and Sui}]{wang2023large}
Peiyi Wang, Lei Li, Liang Chen, Dawei Zhu, Binghuai Lin, Yunbo Cao, Qi~Liu, Tianyu Liu, and Zhifang Sui. 2023.
\newblock Large language models are not fair evaluators.
\newblock \emph{arXiv preprint arXiv:2305.17926}.

\bibitem[{Yao et~al.(2022)Yao, Zhao, Yu, Du, Shafran, Narasimhan, and Cao}]{yao2022react}
Shunyu Yao, Jeffrey Zhao, Dian Yu, Nan Du, Izhak Shafran, Karthik Narasimhan, and Yuan Cao. 2022.
\newblock React: Synergizing reasoning and acting in language models.
\newblock \emph{arXiv preprint arXiv:2210.03629}.

\bibitem[{Yu et~al.(2018)Yu, Zhang, Yang, Yasunaga, Wang, Li, Ma, Li, Yao, Roman et~al.}]{yu2018spider}
Tao Yu, Rui Zhang, Kai Yang, Michihiro Yasunaga, Dongxu Wang, Zifan Li, James Ma, Irene Li, Qingning Yao, Shanelle Roman, et~al. 2018.
\newblock Spider: A large-scale human-labeled dataset for complex and cross-domain semantic parsing and text-to-sql task.
\newblock In \emph{Proceedings of the 2018 Conference on Empirical Methods in Natural Language Processing}, pages 3911--3921.

\end{thebibliography}
\bibliographystyle{acl_natbib}
\end{document}